\documentclass[useAMS,usenatbib]{mn2e}
\usepackage{graphicx}
\usepackage{amsmath,amssymb}
\usepackage{fixltx2e}
\newcommand{\ms}{$\,{\rm M}_\mathrm{\odot}$}

\newcommand{\diff}[2]{\frac{\partial (#1)}{\partial #2}} 

\title[Towards a unified model of stellar rotation]{Towards a unified model of stellar rotation}
\author[A. T. Potter, C. A. Tout \& J. J. Eldridge]{Adrian T. Potter \thanks{E-mail:
apotter@ast.cam.ac.uk} Christopher A. Tout and John J. Eldridge\\
Institute of Astronomy, The Observatories, Madingley Road, Cambridge CB3 0HA}
\begin{document}

\maketitle

\begin{abstract}

The effects of rapid rotation on stellar evolution can be profound. We are now beginning to gather enough data to allow a realistic comparison between different physical models. Two key tests for any theory of stellar rotation are first whether it can match observations of the enrichment of nitrogen, and potentially other elements, in clusters containing rapid rotators and secondly whether it can reproduce the observed broadening of the main sequence in the Hertzsprung-Russel diagram. Models of stellar rotation have been steadily increasing in number and complexity over the past two decades but the lack of data makes it difficult to determine whether such additions actually give a closer reflection of reality. One of the most poorly explored features of stellar rotation models is the treatment of angular momentum transport within convective zones. If we treat the core as having uniform specific angular momentum the angular momentum distribution in the star, for a given surface rotation, is dramatically different from what it is when we assume the star rotates as a solid body. The uniform specific angular momentum also generates strong shears which can drive additional transport of chemical elements close to the boundary of a convection zone. A comparison of different models and their reproduction of observable properties with otherwise identical input physics is essential to properly distinguish between them. We compare detailed grids of stellar evolution tracks  of intermediate and high-mass stars produced using several models for rotation generated with our new stellar rotation code.

\end{abstract}

\section{Introduction}

It has been known for many years that rapid rotation can cause significant changes in the evolution of stars \citep[e.g.][]{Kippenhahn70}. Not only does it cause a broadening of the main sequence in the HR diagram but it also produces enrichment of a number of different elements at the stellar surface \citep[e.g.][]{Hunter09, Frischknecht10}. With new large scale surveys, such as the VLT-FLAMES survey of massive stars, now reaching maturity, the data available for rotating stars are growing rapidly \citep[e.g.][]{Evans05,Evans06}. Any viable model of stellar rotation must be able to match the observed changes in chemical enrichment and structure. The treatment of rotation and its induced chemical mixing in stars has changed dramatically over the past two decades. The model of \citet{Zahn92} has formed the basis for much of this work and many variations from the original have been used during this time to generate different sets of stellar evolution tracks \citep[e.g.][]{Talon97,Meynet00,Maeder03}. Alternative formalisms, such as that of \citet{Heger00}, treat the physical processes very differently. Particular emphasis is often placed on the treatment of meridional circulation. While those models based on Zahn's (1992) treat meridional circulation as an advective process, \citet{Heger00} treat it as a diffusive process. This leads to fundamentally different behaviour. This is just one feature of the model which may lead to significantly different results. Even so, both treatments are used and frequently quoted in the literature for their predictions of the effect of rotation on stellar evolution. 

One of the most poorly explored features of stellar rotation is the treatment of angular momentum transport within convective zones. Current 1D models, treat convective zones as rotating solid bodies. This is not necessarily correct and there are strong reasons to explore alternatives such as uniform specific angular momentum \citep[e.g.][]{Arnett09,Lesaffre10}. This potentially has dramatic consequences for the evolution because a star with uniform specific angular momentum through its convective core has much more angular momentum for a given surface rotation velocity. It also has a strong shear layer at the convective boundary that can drive additional transport of chemical elements.

Different models for stellar rotation have not been compared directly on a common numerical platform alongside otherwise identical input physics before. From the Cambridge stellar evolution code \citep{Eggleton71,Pols95} we have produced a code capable of modelling rotating stars in 1D under the shellular rotation hypothesis of \citet{Zahn92}. The code, \textsc{RoSE} (Rotating Stellar Evolution), can be easily programmed to run with different physics for stellar rotation and can model both radiative and convective zones under a range of different assumptions. This allows us to compare a variety of models for stellar rotation and determine what, if any, observable traits could be used to distinguish between them. We foresee two possibilities; either we can identify clear observational tests to eliminate certain models or the models show no testable difference in which case a simplified model could be formulated to provide the same results.

In section~2 we outline the physical ingredients to \textsc{RoSE} and the different models already implemented. In section~3 we present a comparison of the evolutionary predictions for each model. In section~4 we present our summary and conclusions.

\section{Stellar evolution code}
\label{code}

Here we present the physical ingredients of the numerical code, \textsc{RoSE}, used to implement the models of stellar rotation.

\subsection{Stellar evolution package}

\textsc{RoSE} is based on the Cambridge stellar evolution code, \textsc{STARS}. Originally written by \citet{Eggleton71}, the code has been modified many times over the past thirty years. For details of the last significant update see \citet{Eldridge09}. We now solve the four structure equations, seven chemical equations and now the angular velocity equation in a single, implicit, Newton-Raphson iterative step. We calculate $^1$H, $^4$He, $^3$He, $^{12}$C, $^{14}$N, $^{16}$O and $^{20}$Ne implicitly and 39 other isotopic abundances can be calculated explicitly. The nuclear reaction rates were updated by \citet{Pols95} and \citet{Stancliffe05} and are based on the reaction rates of \citet{Caughlan88}. The opacities were last updated by \citet{Eldridge04} and are based on the OPAL opacities \citep{Iglesias96} at high temperatures and \citet{Ferguson05} at low temperatures. Also included are further corrections for major molecular opacities \citep{Stancliffe08}. The equation of state is that of \citet{Pols95}, convection is treated by mixing-length theory \citep{Bohm58} and a model for convective overshooting \citep{Schroder97} is included. Whilst the code has a number of models for mass loss programmed, we restrict ourselves to the mass-loss rates of \citet{Vink01} for massive stars. These rates apply to non-rotating stars and are modified as explained in section~\ref{massloss}.

\subsection{Structure equations for rotating stars}

The centrifugal force caused by rotation affects the hydrostatic balance of the star, effectively reducing the local gravity. On a surface of constant radius the centrifugal force acts more strongly at the equator than the poles so the distortion of the star depends on co-latitude and our assumption of spherical symmetry is no longer valid. \citet{Tassoul78} showed that, except for stars that are close to critical rotation, the effect of rotational deformation remains axially symmetric. Enhanced mass loss from the surface because of rotation generally keeps stars rotating sufficiently below critical. Even when this is not the case it is only the outer most layers that are affected. In models where the angular momentum distribution in convective regions is uniform the rotation rate may approach critical there but because convective turbulence is already considered to be fully asymmetric, we do not need to consider further non-axial instabilities owing to the rotation. 

We adopt similar adjustments to the stellar structure equations to those described by \citet{Endal76} and \citet{Meynet97}. First we define $S_P$ to be a surface of constant pressure, $P$. $V_P$ is the volume contained within $S_P$ and $r_P$ is the radius of a sphere with volume $V_P=4\pi r_P^3/3$. The equation of continuity is then preserved in its non rotating form,
\begin{equation}
\frac{dm_P}{dr_P}=4\pi r_P^2\rho,
\end{equation}

\noindent where $m_P$ is the mass enclosed within $S_P$ and $\rho$ is the density on the isobar which is assumed to be uniform. As we discuss in section~\ref{rotation} we expect variables and chemical abundances to be uniform across isobars owing to the strong horizontal turbulence caused by the strong density stratification present in stars \citep{Zahn92}. The local gravity vector is
\begin{equation}
\label{geff}
{\boldsymbol g}_{\rm eff}=\left(-\frac{GM}{R^2}+\Omega^2R\sin^2\theta\right) {\boldsymbol e}_{r}+\left(\Omega^2R\sin\theta\cos\theta\right){\boldsymbol e}_{\theta},
\end{equation}

\noindent where $\Omega$ is the local angular velocity. To proceed further we define the average of a quantity over $S_P$ as
\begin{equation}
<q>\equiv\frac{1}{S_P}\oint_{S_P}q d\sigma,
\end{equation}

\noindent where $d\sigma$ is a surface element of $S_P$. Using this notation the equation of hydrostatic equilibrium becomes
\begin{equation}
\frac{dP}{dm_P}=-\frac{Gm_P}{4\pi r_P^4}f_P,
\end{equation}

\noindent where 
\begin{equation}
f_P=\frac{4\pi r_P^4}{Gm_PS_P}<g_{\rm eff}^{-1}>^{-1}
\end{equation}

\noindent and $g_{\rm eff}\equiv |{\boldsymbol g}_{\rm eff}|$. Hence with the new definition of variables we can retain the same 1D hydrostatic equilibrium equation modified by a factor of $f_P$ which tends to unity for no rotation.

The equation for radiative equilibrium is similarly modified to
\begin{equation}
\frac{d \ln T}{d \ln P}=\frac{3\kappa P L_P}{16 \pi acGm_PT^4}\frac{f_T}{f_P},
\end{equation}

\noindent where $L_P$ is the total energy flux through $S_P$, $P$ is the pressure, $T$ is the temperature, $\kappa$ is the opacity, $a$ is the radiation constant, $c$ is the speed of light, $G$ is the gravitational constant and
\begin{equation}
f_T\equiv\left(\frac{4\pi r_P^2}{S_P}\right)\left(<g_{\rm eff}><g_{\rm eff}^{-1}>\right)^{-1}.
\end{equation}

\noindent Again, the non-rotating equation for stellar evolution has been preserved except for the multiplication by $f_T/f_P$. Of the two factors, $f_P$ deviates further from unity for a given rotation than $f_T$. Additional secondary effects of the reduced gravity must be taken into account when calculating quantities such as the pressure scale height and Brunt--V\"ais\"al\"a frequency. For the remainder of the paper we drop the subscript $P$.

\subsection{Meridional circulation}

The amount of thermal flux, $F$ through a point in a star behaves as $F\propto g_{\rm eff}(\theta)$. As in equation~(\ref{geff}), $g_{\rm eff}$ is strongly dependent on co-latitude and so the radiative flux is greater at the poles than at the equator. This leads to a global thermal imbalance that drives a meridional circulation. The presence of meridional circulation has been considered for nearly a century since \citet{VonZeipel24}. In the past it has commonly been approximated by Eddington-Sweet circulation \citep{Sweet50}. \citet{Zahn92} proposed an alternative but similar treatment of the meridional circulation based on energy conservation along isobars and this is the formulation we are currently using. In spherical polar coordinates the circulation takes the form
\begin{equation}
{\boldsymbol U}=U(r)P_2(\cos\theta){\boldsymbol e}_r+V(r)\frac{dP_2(\cos\theta)}{d\theta}{\boldsymbol e}_{\theta},
\end{equation}

\noindent where $U$ and $V$ are linked by the continuity equation
\begin{equation}
V=\frac{1}{6\rho r}\frac{d}{dr}(\rho r^2U).
\end{equation}

\noindent and $P_2$ is the second Legendre polynomial $P_2(x)=\frac{1}{2}(3x^2-1)$. We currently use an approximate form for the meridional circulation by \citet{Maeder98}
\begin{equation}
\begin{split}
U=C_0\frac{L}{m_{\rm eff}g_{\rm eff}}\frac{P}{C_P\rho T}\frac{1}{\nabla_{\rm{ad}}-\nabla+\nabla_{\mu}}\\\left(1-\frac{\epsilon}{\epsilon_m}-\frac{\Omega^2}{2\pi G\rho}\right)\left(\frac{4\Omega^2r^3}{3Gm}\right),
\end{split}
\end{equation}

\noindent where $m_{\rm eff}=m\left(1-\frac{\Omega^2}{2\pi G\rho}\right)$, $\epsilon=E_{\rm nuc}+E_{\rm grav}$, the total local energy emission, $\epsilon_m=L/m$, $C_P$ is the specific heat capacity at constant pressure, $\nabla$ is the radiative temperature gradient, $\nabla_{\rm{ad}}$ is the adiabatic temperature gradient and $\nabla_{\mu}$ is the mean molecular weight gradient. For simplicity, the thermodynamic ratio $\frac{\gamma}{\delta}$ used by \citet{Maeder98} is set to unity. This is the correct for a perfect gas. We have also approximated the factor $\tilde{g}/g$ of \citet{Zahn92} by $\frac{4\Omega^2r^3}{3Gm}$. This is a suitable approximation throughout most of the star. The constant $C_0$ is included for aid of calibration and is discussed further in section \ref{testmodels}.

The same formulation for the meridional circulation is used in all the test models except those that use the formalism of \citet{Heger00} where the circulation is treated as a diffusive rather than an advective process and the characteristic circulation velocity is calculated from Eddington-Sweet circulation.

\subsection{Mass loss with rotation}
\label{massloss}

Observational evidence for enhanced mass loss is mixed \citep[see][]{Vardya85, Nieuwenhuijzen88} but theoretically near-critical rotation must drive additional mass loss to remove angular momentum and prevent the surface of the star from rotating super-critically \citep{Friend86}. We use the enhanced mass-loss rate of \citet{Langer98}
\begin{equation}
\dot{M}=\dot{M}_{\Omega=0}\left(\frac{1}{1-\frac{\Omega}{\Omega_{\rm crit}}}\right)^{\xi},
\end{equation}

\noindent where we take $\xi=0.45$ and $\Omega_{\rm crit}=\sqrt{\frac{2GM}{3R^3}}$. A more complete discussion of the critical rotation rates of stars is given by \citet{Maeder00} and \citet{Georgy11}.

\subsection{Rotation in convective zones}

Current 1D models of stellar evolution generally assume that convective zones are kept in solid body rotation. This may be caused by strong magnetic fields induced by dynamo action \citep{Spruit99} but there is no conclusive evidence that real fields generated by this mechanism are strong enough to enforce solid body rotation. Certainly in the Sun we see latitudinal variations in the angular velocity throughout the outer convective layer \citep{Schou98}. Standard mixing length theory suggests that a rising fluid parcel should conserve its angular momentum before mixing it with the surrounding material after rising a certain distance. This would lead to uniform specific angular momentum rather than solid body rotation. This is supported by a recent MLT-based closure model for rotating stars \citep{Lesaffre10} and by 3D hydrodynamic simulations \citep{Arnett09}. In reality magnetic fields are likely to play some part in the transport of angular momentum but it is uncertain whether these are strong enough to affect the hydrodynamics. The asymptotic behaviour of the rotation profile in convective zones could have a profound effect on the evolution of the star, first because the total angular momentum content of a star for a given surface rotation increases dramatically for uniform specific angular momentum and secondly because uniform angular momentum in the convective zone produces a layer of strong shear at the boundary with the radiative zone and this drives additional chemical mixing. To explore the different possible behaviours we have introduced, in \textsc{RoSE}, the capacity to vary the distribution of angular momentum in convective zones as discussed in section~\ref{rotation}.

\subsection{Angular momentum transport}
\label{rotation}

Differential rotation is expected to arise in stars because of hydrostatic structural evolution, mass loss and meridional circulation. Because of this stars are subject to a number of local hydrodynamic instabilities. These are expected to cause diffusion of radial and latitudinal variations in the angular velocity in order to bring the star back to solid body rotation, its lowest energy state. This occurs with characteristic diffusion coefficients $D_{\rm shear}$ and $D_{\rm h}$ respectively. \citet{Zahn92} proposed that, because of the strong stratification present in massive stars, the turbulent mixing caused by these instabilities is much stronger horizontally than vertically ($D_{\rm h}\gg D_{\rm shear}$). This leads to a situation where the angular velocity variations along isobars are negligible compared to vertical variations. Furthermore, all other state variables are assumed to be roughly constant over isobars and the mixing produces horizontal chemical homogeneity. This is referred to as shellular rotation, where we describe the angular velocity by $\Omega=\Omega(r)$.

Taking into account all of the processes described in section~\ref{code} we use the evolution equation for the angular velocity \citep{Zahn92}
\begin{equation}
\begin{split}
\label{main1}
\diff{r^2\Omega}{t}=\frac{1}{5\rho r^2}\diff{\rho r^4\Omega U}{r}+\frac{1}{\rho r^2}\frac{\partial}{\partial r}\left(\rho {D_{\rm{shear}} r^4\frac{\partial\Omega}{\partial r}}\right)\\+\frac{1}{\rho r^2}\frac{\partial}{\partial r}\left(\rho {D_{\rm{conv}} r^{(2+n)}\frac{\partial r^{(2-n)}\Omega}{\partial r}}\right)
\end{split}
\end{equation}

\noindent and for the chemical evolution
\begin{equation}
\label{main2}
\frac{\partial c_i}{\partial t}=\frac{1}{r^2}\frac{\partial}{\partial r}\left(\left(D_{\rm{shear}}+D_{\rm{eff}}+D_{\Omega=0}\right)r^2\frac{\partial c_i}{\partial r}\right),
\end{equation}

\noindent where $c_i$ is the abundance of species $i$. The diffusion coefficient $D_{\rm conv}$ is non-zero only in convective zones and the $D_{\rm shear}$ and $D_{\rm eff}$ are non-zero only in radiative zones. The parameter $n$ sets the steady state specific angular momentum distribution in convective zones, $n=2$ corresponds to solid body rotation and $n=0$ corresponds to uniform specific angular momentum. The coefficient $D_{\rm eff}$ describes the effective diffusion of chemical elements because of the interaction between horizontal diffusion and meridional circulation so
\begin{equation}
D_{\rm eff}=\frac{|rU|^2}{30 D_{\rm h}}.
\end{equation}

The main differences between models \citep[e.g.]{Talon97,Heger00,Meynet00,Maeder03} are the treatment of the meridional circulation, $U$, the diffusion coefficients, $D_{\rm shear}$, $D_{\rm eff}$ and $D_{\rm conv}$, and the steady power-law distribution of angular momentum in convective zones determined by $n$. We describe many of these models which we have implemented with \textsc{RoSE} in section~\ref{testmodels}.
\subsection{Test cases}
\label{testmodels}

In this paper we consider a number of common models for comparison. The details of the models are summarized in Table \ref{modeltable}. Unless otherwise stated, the metallicity is taken to be solar ($Z=0.02$). For each model we compute the stellar evolution for a range of masses between $3$\ms\ and $100$\ms\ and initial equatorial surface rotation velocities between $0$ and $600\,{\rm km\,s^{-1}}$, except where the initial surface rotation rate would be super-critical.

\begin{table*}
\caption{
\label{modeltable}
The diffusion coefficients used by each of the different models.}
\begin{tabular}{ccccc}
\hline
Case & $n$ & $D_{\rm shear}$ & $D_{\rm h}$ & $D_{\rm conv}$\\
\hline
1 & 2 & \citet{Talon97} & \citet{Maeder03} & $D_{\rm mlt}$\\
2 & 2 & \citet{Heger00} & N/A &  $D_{\rm mlt}$\\
3 & 2 & \citet{Zahn92} & \citet{Zahn92} & $D_{\rm mlt}$\\
4 & 0 & \citet{Talon97} & \citet{Maeder03} & $D_{\rm mlt}$\\
5 & 0 & \citet{Talon97} & \citet{Maeder03} & $10^{10}{\rm cm^2\,s^{-1}}$\\
6 & 0 & \citet{Talon97} & \citet{Maeder03} & $10^{14}{\rm cm^2\,s^{-1}}$\\
\hline
\end{tabular}
\end{table*}

\subsubsection{Case 1:}

Here we use the formulation for $D_{\rm shear}$ of \citet{Talon97},
\begin{equation}
D_{\rm shear}=\frac{2 Ri_{\rm c}\left(r\frac{d\Omega}{dr}\right)^2}{N_T^2/(K+D_{\rm h})+N_{\mu}^2/D_{\rm h}},
\end{equation}

\noindent where $Ri_{\rm c}=(0.8836)^2/2$ is the critical Richardson number which we have taken to be the same as did \citet{Maeder03}.
\begin{equation}
N_T^2=-\frac{g_{\rm eff}}{H_P}\left(\frac{\partial\ln\rho}{\partial \ln T}\right)_{\!\!P,\mu}\left(\nabla_{\rm ad}-\nabla\right)
\end{equation}
\noindent and
\begin{equation}
N_{\mu}^2=\frac{g_{\rm eff}}{H_P}\left(\frac{\partial\ln\rho}{\partial\ln\mu}\right)_{\!\!P,T}\frac{d \ln\mu}{d \ln P}.
\end{equation}

\noindent Following \citet{Maeder03} we take
\begin{equation}
D_{\rm h}=0.134 r\left(r\Omega V\left[2V-\alpha U\right]\right)^{\frac{1}{3}}{,}
\end{equation} 

\noindent where
\begin{equation}
\alpha=\frac{1}{2}\frac{d(r^2\Omega)}{dr}.
\end{equation}

\noindent The differential equations derived by \citet{Zahn92} are fourth order in space. Our model differs in that third order derivatives and above cannot be reliably computed and must be ignored. The constant $C_0$ is included as a means of calibrating the model in light of this difference. Because our ultimate intention is to compare these models to data from the VLT-FLAMES survey of massive stars, we have chosen $C_0$ so that we reproduce the terminal-age main-sequence  (TAMS) nitrogen enrichment of a $40$\ms$\,$ star initially rotating at $270{\,\rm km\,s^{-1}}$ with galactic composition given by \citet{Brott10}. This gives $C_0=0.003$. Whilst this is admittedly quite small, it is important to note that the non-linearity of the angular momentum transport equation means that a small change in the amount of diffusion corresponds to quite a large change in $C_0$.  We could have similarly adjusted the magnitude of $D_{\rm shear}$ in case~1 to give the desired degree of nitrogen enrichment. However, attempting to reduce the diffusion coefficient results in more shear which drives the diffusion back up so $D_{\rm shear}$ would need to be reduced by a large factor to give the desired effect. This would have the further consequence that the meridional circulation would have a much stronger effect on the system and produce stars with a very high degree of differential rotation between the core and the envelope. Hence we have chosen to adopt this method of calibration as the most physically reasonable.

\noindent The diffusion of angular momentum in convective zones is determined by the characteristic eddy viscosity from mixing length theory such that $D_{\rm conv}=D_{\rm mlt}=\frac{1}{3}v_{\rm mlt}l_{\rm mlt}$. This model takes $n=2$.

\subsubsection{Case 2:}

This is the model of \citet{Heger00}. In this case we set $U=0$ because circulation is treated as a diffusive process. The details of the various diffusion coefficients are extensive so we refer the reader to the original paper. With their notation the diffusion coefficients are
\begin{equation}
D_{\rm shear}=D_{\rm sem}+D_{\rm DSI}+D_{\rm SHI}+D_{\rm SSI}+D_{\rm ES}+D_{\rm GSF}
\end{equation}

\noindent and
\begin{equation}
D_{\rm eff}=(f_c-1)(D_{\rm DSI}+D_{\rm SHI}+D_{\rm SSI}+D_{\rm ES}+D_{\rm GSF}),
\end{equation}

\noindent where each of the $D_i$ corresponds to a different hydrodynamical instability. We note that our notation differs slightly from the original paper. The diffusion coefficients $\nu$ and $D$ used by \citet{Heger00} are equivalent to $D_{\rm shear}$ and $D_{\rm shear}+D_{\rm eff}$ respectively. \citet{Heger00} take $f_{\rm c}=1/30$ which is what we use here. The parameter $f_{\mu}$ used in \citet{Heger00} is taken to be zero. Mean molecular weight gradients play an important part in chemical mixing near the core however we have performed a number of test runs with $f_{\mu}=0.05$. Although there were some differences, they were not significant and could be largely masked by modifying the other free parameters associated with this case. The model differs from the original formulation of \citet{Heger00} in that we are unable to use {\it STARS} to consistently compute non-local quantities such as the spatial extent of instability regions used in some of the expressions for the various diffusion coefficients. We do not expect this to have a significant effect on the results since $D_{\rm ES}$ dominates the total diffusion coefficient and its limiting length scale is the pressure scale height rather than the extent of the unstable region.

This model is calibrated by modifying the dominant diffusion coefficient for transport owing to meridional circulation, $D_{\rm ES}$, by a constant of order unity to give the same TAMS nitrogen enrichment as case~1 for a $20$\ms\ star with initial surface angular velocity of $300\,{\rm km\,s^{-1}}$. This model has $n=2$ and $D_{\rm conv}=D_{\rm mlt}$.

\subsubsection{Case 3:}

This is a reproduction of the original model of \cite{Zahn92} and is included as a baseline to highlight the differences in predictions of stellar rotation from the original model. In this model
\begin{equation}
D_{\rm shear}=\nu_{\rm v,h}+\nu_{\rm v,v},
\end{equation}

\noindent where
\begin{equation}
\nu_{\rm v,v}=\frac{8 C_1 Ri_{\rm c}}{45}K\left(\frac{r}{N_T^2}\frac{d\Omega}{dr}\right)^2,
\end{equation}
\begin{equation}
\nu_{\rm v,h}=\frac{1}{10}\left(\frac{\Omega}{N_T^2}\right)\left(\frac{K}{D_{\rm h}}\right)^{\frac{1}{2}}r|2V-\alpha U|,
\end{equation}

\noindent $K$ is the thermal diffusivity and $\alpha$ is the same as in case~1. In this model
\begin{equation}
D_{\rm h}=C_2r|2V-\alpha U|,
\end{equation}

\noindent and $C_1$ and $C_2$ are constants used for calibration. We constrain $C_1$ and $C_2$ by matching as closely as possible the TAMS nitrogen enrichment and luminosity of a $20$\ms$\,$ case-1 star with initial surface rotation of $300\,{\rm km\,s^{-1}}$. We find that $C_1=0.019$ which is surprisingly small. This is because this model does not take into account mean molecular weight gradients and this leads to far more mixing between the convective core and the radiative envelope than in case~1. The TAMS luminosity is always greater than case~1 and so we minimize it with respect to $C_2$ so $C_2\to\infty$. This is realised by setting $D_{\rm eff}=0$ and is the case when the horizontal diffusion completely dominates over the meridional circulation and so is consistent with our assumption of shellular rotation.

Apart from ignoring mean molecular weight gradients, the main objection to this model is that in the formulation of $D_{\rm h}$ we assume that, if the horizontal variation in the angular velocity along isobars takes the form $\tilde{\Omega}=\Omega_2(r)P_2(\cos\theta)$, then $\Omega_2(r)/\Omega(r)$ is constant and this is not physically justified. Again we set $n=2$ and $D_{\rm conv}=D_{\rm mlt}$.

\subsubsection{Cases 4, 5 and 6:}

For these cases we use the same $D_{\rm shear}$ and $D_{\rm h}$ as in case~1 but we set $n=0$ to produce uniform specific angular momentum through the convective zones and test the effect of varying the convective diffusion coefficient $D_{\rm conv}$.
\begin{equation}
D_{\rm conv}=
\begin{cases}
D_{\rm mlt}& {\rm case\; 4,} \\10^{10}\,{\rm cm^2\,s^{-1}} & {\rm case\; 5,}\\10^{14}\,{\rm cm^2\,s^{-1}} & {\rm case\; 6.}
\end{cases}
\end{equation}

\subsubsection{Metallicity variation}

We have also calculated the evolution in cases 1, 2 and 3 for the same masses and velocities but with $Z=0.001$. We shall represent these cases with a superscript {\it Z} (e.g. Case ${\rm 1}^Z$ is the low metallicity analogue of Case~1).

\section{Results}

Whilst there are many potential observables which may be used to distinguish between different models, it is important to choose the ones that are most easily compared with observational data. From our stellar evolution calculations we find a number of important differences between the test cases.

\subsection{Evolution of a $20$\ms\ star in cases 1, 2 and 3}

\begin{figure*}
\begin{center}
\includegraphics[width=0.85\textwidth]{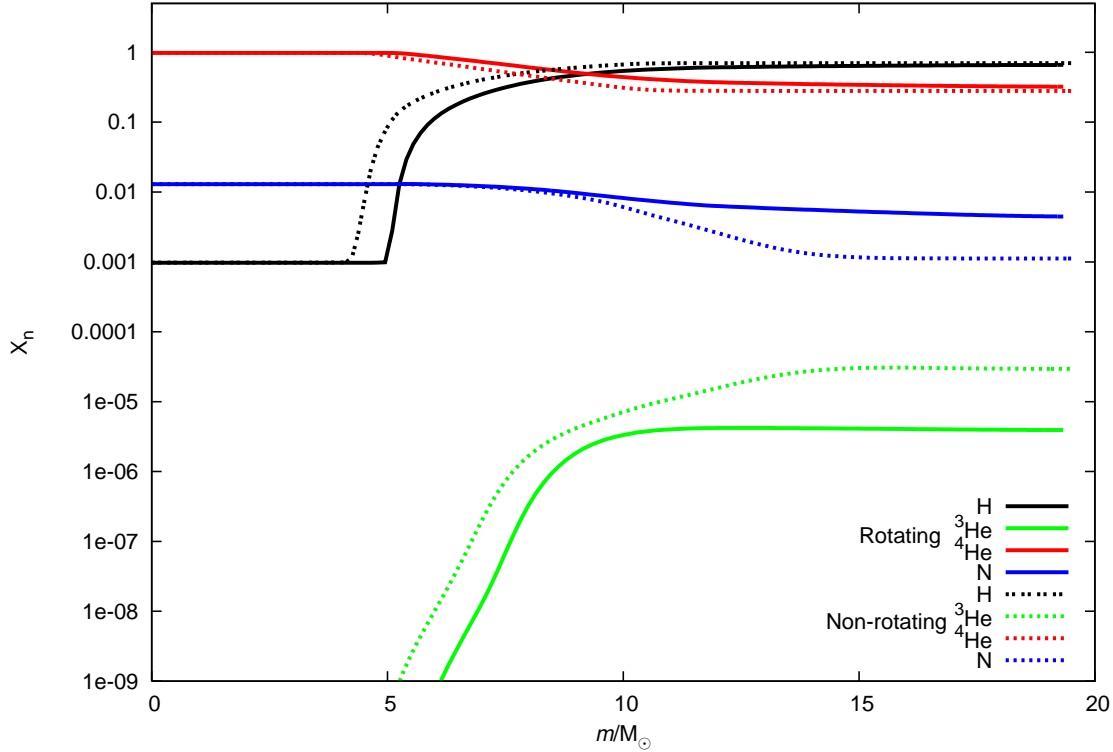}
\end{center}
\caption{Terminal-age main sequence composition of  $20$\ms\ case-1 stars. The solid lines are for a star initially rotating at $300\,{\rm km\,s^{-1}}$ and the dashed lines are for a non-rotating star. Both stars have central hydrogen abundance $X_{\rm H}=10^{-3}$. Note that rotational mixing results in a larger core and mixing of helium and nitrogen throughout the radiative envelope.}
\label{abundances}
\end{figure*}

\begin{figure*}
\begin{center}
\includegraphics[width=0.48\textwidth]{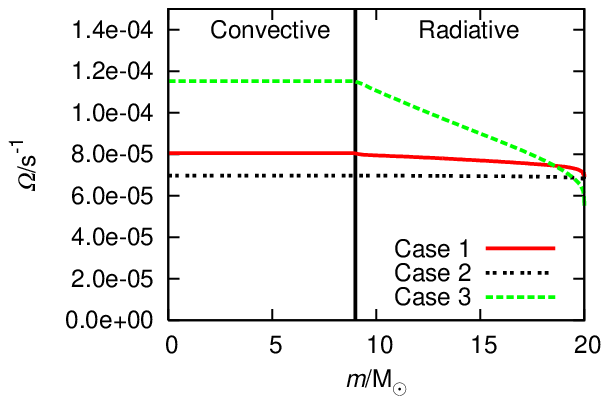} \includegraphics[width=0.48\textwidth]{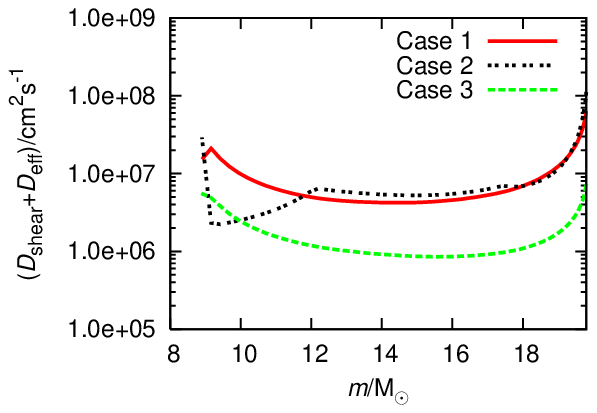}
\end{center}
\caption{Zero-age main sequence properties of a $20$\ms\ star initially rotating at $300\,{\rm km\,s^{-1}}$. The left panel shows the angular velocity through the star and the right panel shows the diffusion coefficient for chemical transport through the radiative envelope}
\label{diff}

\end{figure*}

First it is helpful to briefly examine the internal evolution that occurs. We consider here the main-sequence evolution of a $20$\ms\ star with initial surface rotation of $300\,{\rm km\,s^{-1}}$ for cases 1, 2 and 3. Although the centrifugal force causes some change in a star's structure, its evolution is strongly affected by changes in the chemical composition. Fig \ref{abundances} shows how the composition of the rotating $20$\ms\ case-1 star differs from a non-rotating $20$\ms\ star at the end of the main sequence. The difference in the rotation-induced mixing produces the variation in results between the various cases. Fig. \ref{diff} shows the angular velocity profile and the diffusion coefficient for vertical angular momentum transport in radiative zones predicted by each of cases 1, 2 and 3 at the zero-age main sequence (ZAMS). Note that, even though the stars have the same surface rotation, their core rotation and hence total angular momentum content can vary significantly between models

\begin{figure*}
\begin{center}
\includegraphics[width=0.48\textwidth]{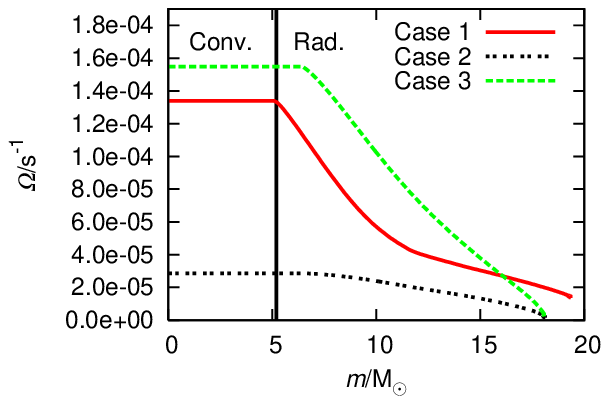} \includegraphics[width=0.48\textwidth]{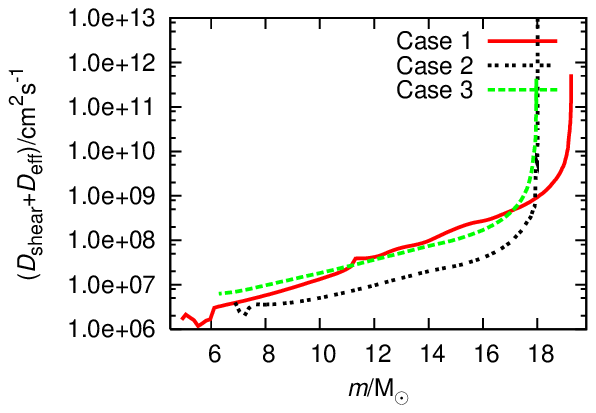}
\end{center}
\caption{Terminal-age main sequence properties of a $20$\ms\ star initially rotating at $300\,{\rm km\,s^{-1}}$. The left panel shows the angular velocity through the star and the right panel shows the diffusion coefficient for chemical transport through the radiative envelope. The vertical line of the left panel separates the convective and radiative zones of the case-1 star.}
\label{diff2}
\end{figure*}

Despite their similar treatments, cases~1 and 3 have quite different initial rotation profiles. This is largely due to our choice of calibration. Because in case~3 we ignore mean molecular weight gradients, the overall efficiency of mixing must be reduced to match our calibration criterion (section \ref{testmodels}). This means that shear diffusion is much weaker relative to advection and so a profile with more differential rotation results. Had we chosen to calibrate the mixing by reducing $C_0$ instead of $C_1$ we would have found the opposite effect. This highlights one possible pitfall of including multiple free parameters within a given system.

 Case~2 is dominated by diffusion because of the diffusive treatment of meridional circulation. Recall that meridional circulation is treated advectively in cases 1 and 3 so is not included in the diffusion coefficient. In fact it is responsible for production of the shear at the ZAMS despite turbulence trying to restore solid body rotation. Because there is no perturbation to the rotation at the start of the main sequence, the star in case~2 rotates as a solid body. As the star evolves and mass is lost from its surface the solid body rotation is disturbed. Even so, because of the strong diffusion, case~2 stars never deviate far from solid body rotation as can be seen in Fig. \ref{diff2}. We note that case-1 stars reach the end of the main sequence with a higher mass than those in case~2 or case~3. This is because case-2 and case-3 stars have a longer main-sequence life owing to more efficient mixing at the core--envelope boundary. This allows hydrogen to be mixed into the core more rapidly than in case~1. This also leads to larger core masses in cases~2 and~3 compared to case~1.

We see in Fig. \ref{diff} that the predicted diffusion coefficients for cases~1 and~2 are similar throughout most of the envelope. By the TAMS, the diffusion predicted by case~2 is significantly lower than the other two cases. This is possibly because rising shear owing to rapid hydrostatic evolution at the TAMS causes the diffusion in cases~1 and 3 to increase while in case~2 diffusion is dominated instead by the circulation. Unsurprisingly, the diffusion coefficient in case~3 is similar in form to case~1 but significantly smaller, a result of our choice of $C_1$. We note however that the diffusion coefficient predicted by the two cases is very similar by the end of the main sequence. Also, the diffusion coefficient at the core-envelope boundary at the end of the main sequence in case~1 is around an order of magnitude lower than in case~3. This is partially due to the core of the case-3 star rotating faster than in case~1 but is mostly due to the inclusion of the mean molecular weight gradient in the formulation for case~1. \citet{Frischknecht10} (who use a model very similar to our case~1) predict a much greater decline in the mixing near the core but we have been unable to reproduce this. It is likely that the difference is a result of the inhibiting effect of the mean molecular weight gradient on the rotational mixing. Whilst it is included in our models, the results of \citet{Frischknecht10} suggest that towards the end of the main-sequence its effect covers a much larger proportion of the radiative envelope than in our models.

\subsection{Effect on HR diagram}

\begin{figure*}
\begin{center}
\includegraphics[width=0.85\textwidth]{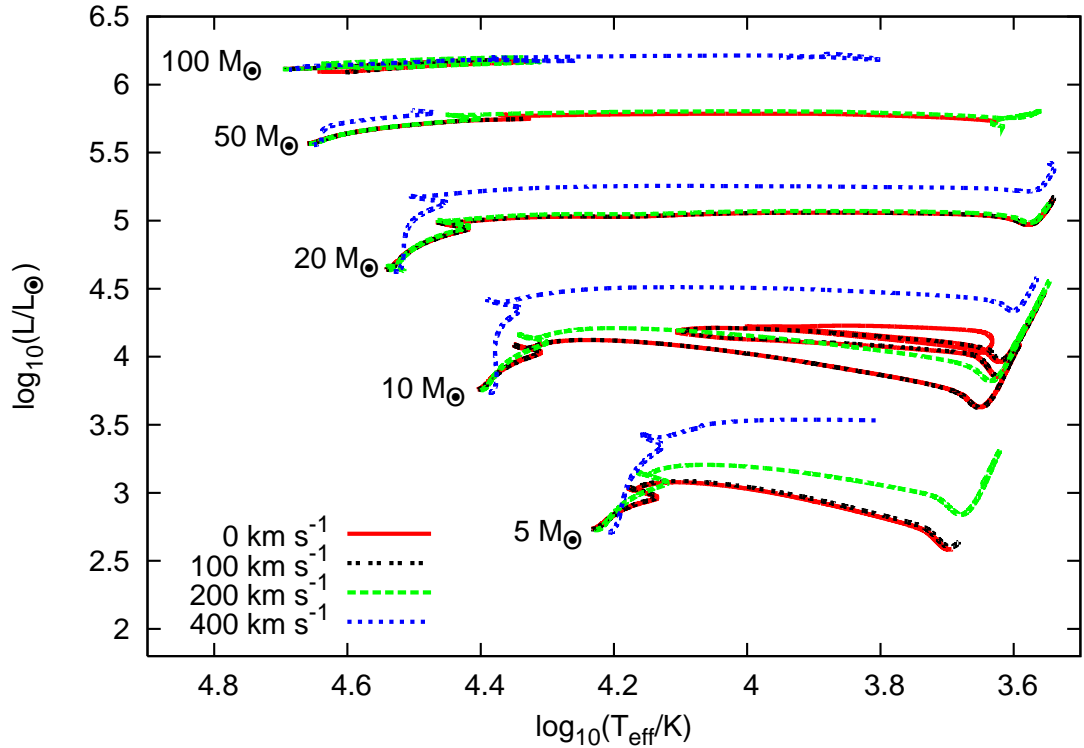}
\end{center}
\caption{Stellar evolution tracks for stars between $5$\ms\ and $100$\ms\ calculated with \textsc{RoSE} for case~1 with surface velocities of $0$ to $400\,{\rm km\,s^{-1}}$. The tracks for cases 4, 5 and 6, where the angular momentum transport in convective zones is varied, are almost indistinguishable from each other but produce more luminous stars than in case~1. There is significant difference between the predictions made for cases 2 and 3, where the angular momentum transport in radiative zones is varied.}
\label{HR}
\end{figure*}

As expected, the effects of rotation on the structure and chemical evolution of each star are significant across the HR diagram. We have plotted the case~1 models in Fig \ref{HR}. Centrifugal forces cause the star to expand making it dimmer and redder. However, because of the additional chemical mixing, more hydrogen is made available during the main sequence so, by the main-sequence turn off, rotating stars are generally more luminous than similar mass non-rotating stars. This effect becomes more pronounced at higher rotation rates and is most apparent for stars with masses up to $20$\ms.

\begin{figure*}
\begin{center}
\includegraphics[width=0.85\textwidth]{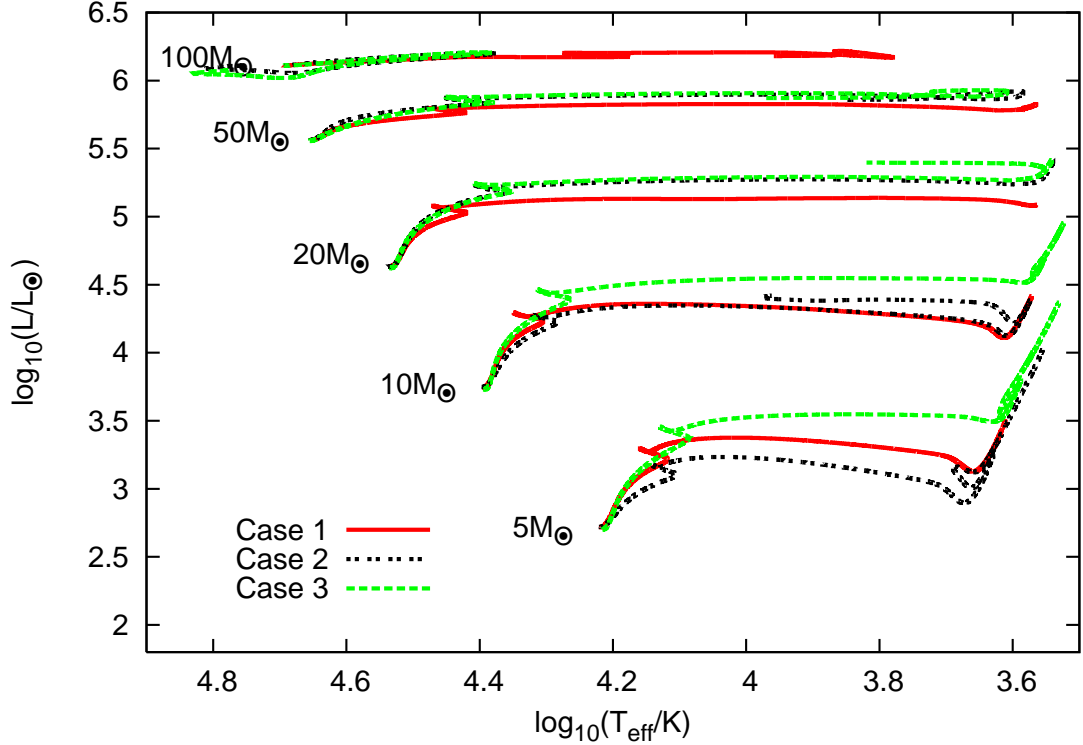}
\end{center}
\caption{Stellar evolution tracks for stars between $5$\ms\ and $100$\ms\ with initial surface rotation of $300\,\rm{km}\,\rm{s}^{-1}$ for cases 1, 2 and 3. The model of \citet{Heger00} predicts a greater enhancement and higher surface temperatures compared to that based on \citet{Maeder03} for stars more massive than $10$\ms.}
\label{HRrad1}
\end{figure*}

There are clear differences between the predicted evolution of rotating stars in each of the three cases. Fig. \ref{HRrad1} shows the evolution of five different stellar masses in the HR diagram for cases 1, 2 and 3 with an initial surface rotation of $300\,\rm{km}\,\rm{s}^{-1}$. These cases are the three different models of rotational mixing in radiative zones at solar metallicity. Rotating case-1 stars appear to be the most luminous at low masses but least luminous at high masses. Case-2 stars are also consistently cooler than their case-1 and case-3 equivalents except below $10$\ms. This is because the strength of rotation-induced mixing increases rapidly with mass in case-2 stars unlike in cases 1 and 3 where the difference is more modest.

Although there is apparently a large difference between the three models in the HR diagram, to distinguish between them from stellar populations requires either a very large sample or accurate independent measurements of stellar masses and rotation velocities. Both of these are very challenging but, with the advent of large scale surveys, the former is quickly becoming a possibility.

\subsection{Nitrogen enrichment}

Currently the key test for any model of stellar rotation is how well it reproduces the spread of data in a Hunter diagram \citep{Hunter09}. Hunter plotted nitrogen enrichment against surface rotation. Large scale surveys, such as the VLT-FLAMES Tarantula survey \citep{Evans10}, will greatly increase the data available for surface rotation velocities and surface abundances over the coming decade. Thus, if different models can be distinguished in a Hunter diagram, this would form a key test for stellar rotation models.

\begin{figure*}
\centering
\includegraphics[width=0.47\textwidth]{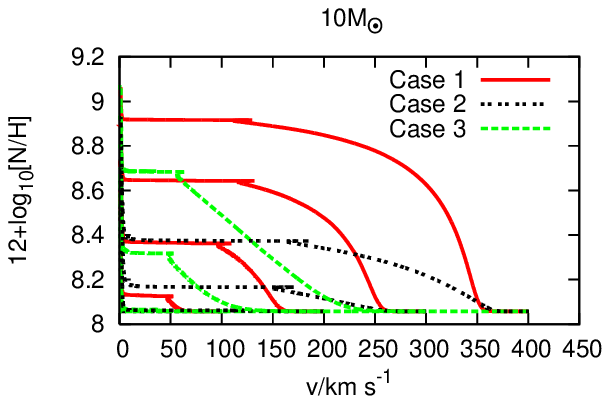} \includegraphics[width=0.47\textwidth]{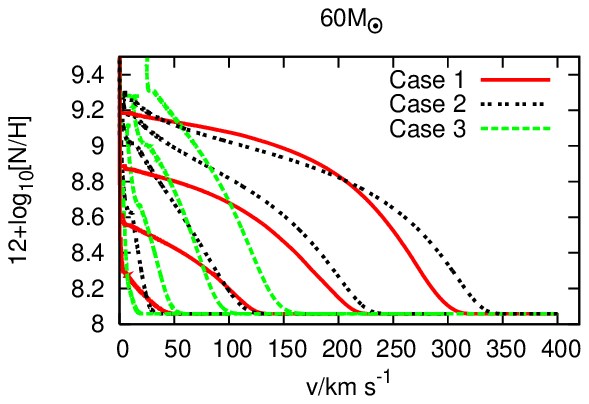}
\caption{Nitrogen enrichment (by number of nuclei) variation with initial surface rotation for cases 1, 2 and 3. Stars start on the ZAMS with low nitrogen abundances and high velocities and evolve to higher abundances and lower velocities during the main sequence. The left panel is for $10$\ms\ stars and the right panel is for $60$\ms\ stars. Note the different scales for each panel. As expected the enrichment is much greater for more massive stars. Case-3 stars spin down more before any enrichment occurs. They give greater enrichment than either of the other two models at high masses but significantly less than case~1 for low masses. Cases~1 and 2 enrich to a similar degree for the high-mass stars but case~2 exhibits significantly less enrichment for low-mass stars than case~1.}
\label{Nsol}
\end{figure*}

In Fig. \ref{Nsol} we plot our theoretical Hunter diagram for $10$\ms\ and $60$\ms\ stars at different initial surface rotation velocities with the different radiative-zone models at solar metallicity. Each star begins at the bottom of the plot with the same nitrogen abundance but different initial surface velocities. There is an initial period where the star spins down before any enrichment has occurred. During this time the star moves straight to the left of the plot. Eventually the surface nitrogen abundance begins to increase because of rotation induced mixing from the burning region. At the same time the star continues to spin down because of mass loss and structural evolution. The net effect is that the star moves towards the upper left-hand region of the plot. At the end of the main sequence the star expands and rapidly spins down. This appears as a near-horizontal line as the star moves rapidly towards the left-hand edge of the plot. Some further enrichment may occur during the giant phase.

The evolution of the surface abundances is very model dependent. Case~3, which is based on the early model of \citet{Zahn92}, gives more nitrogen enrichment than case~1 at high masses ($M\geq 60$\ms) but significantly less at low masses ($M\leq 10$\ms). Most notably though, the case-3 stars spin down to a far greater degree before enrichment occurs. We attribute this to the neglect of mean molecular weight gradients. Mixing near the core is more efficient in case~3 but, due to the overall calibration, is weaker near the surface.

For case~2 the amount of mixing in lower-mass stars is much less than for both cases 1 and 3. By comparison the enrichment of case-2 $60$\ms\ stars is greater than in the other two cases, particularly for slower rotators. This mass dependent behaviour of the rotating models could provide important clues to distinguish between the models.

At solar metallicity, owing to the enhanced mass loss, massive stars spin down before the end of the main sequence so, in this case, we would not observe the absence of the moderately rotating, highly enriched stars seen at low metallicity \citep{Hunter09}. Observations of multiple clusters at different ages at this metallicity would be a good test for rotating stellar models because the evolution across the Hunter diagram is significantly different in each case~even when they are calibrated to give the same level of enrichment at the end of the main sequence.

\subsection{Helium-3 enrichment}

Apart from the enrichment of nitrogen, rotation can have a large effect on the evolution of other elements. Changes in the carbon and oxygen abundances in rotating stars predicted from models have been considered but the accuracy of the data prohibits any strong conclusions from being made. \citet{Frischknecht10} discuss the effect rotation may have on the surface abundances of light elements. We consider here the evolution of the surface abundance of $^3$He. A similar analysis could be performed for other elements such as lithium and boron. The changes in the overall abundance of $^3$He because of rotation could partially explain the discrepancy between the predicted abundances produced by stars and the lack of enrichment of the ISM compared to levels predicted by primordial nucleosynthesis \citep{Dearborn86,Hata95,Dearborn96}. This has been explained in the past by thermohaline mixing \citep{Stancliffe10} but, as our results show, the surface $^3$He abundance is strongly affected by rotation and so it is likely to make at least some contribution to this effect. We leave the issue of whether the total production increases or decreases over the stellar lifetime to future work. In either case, the evolution of helium-3 with respect to the surface rotation is very different between alternative models and so, as for nitrogen enrichment, this would form a useful means to compare stellar rotation models. Unlike nitrogen, helium-3 enrichment is stronger at low masses and so provides a greater number of candidate stars for observations.

\begin{figure*}
\centering
\includegraphics[width=0.47\textwidth]{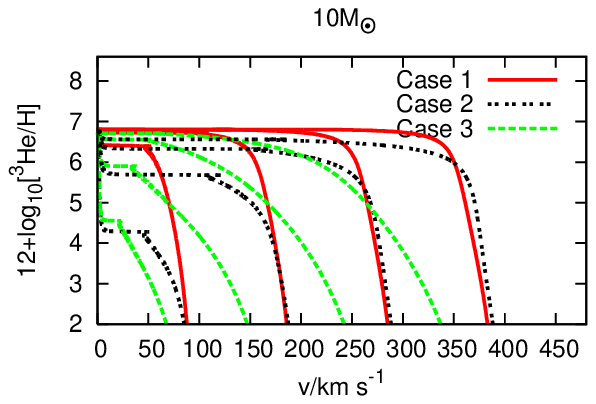} \includegraphics[width=0.47\textwidth]{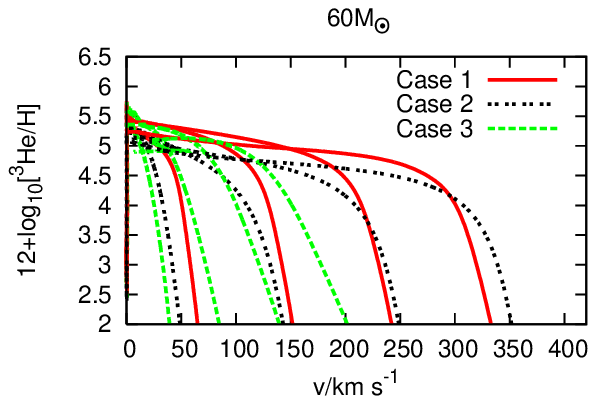}
\caption{Helium-3 enrichment variation with initial surface rotation for cases 1, 2 and 3. The left panel is for $10$\ms\ stars and the right panel is for $60$\ms\ stars. At high masses all three cases show a similar degree of enrichment though there is some variation in the amount. Case-3 stars spin down more before enrichment occurs and so lie to the left of the other two cases. At low masses, both case-2 and case-3 stars are less enriched at the TAMS than case-1 stars. This is especially true for slow rotators.}
\label{He3sol}
\end{figure*}

Fig. \ref{He3sol} shows the helium-3 enrichment for $10$\ms\ and $60$\ms\ stars of varying initial surface velocities for each of the different radiative zone models at solar metallicity. As for nitrogen, all three cases show comparable amounts of enrichment in high-mass stars. The amount of enrichment at the end of the main sequence is the same in each case~but case-3 stars are slightly more enriched at all rotation rates. Case-3 stars spin down much more before enrichment occurs so the paths for these stars lie to the left of case~1 and case~2 stars but the amount of enrichment at the end of the main sequence is comparable, though slightly higher than, the other two cases.

The difference between the test cases is far greater at lower masses. Both case-2 and case-3 stars show substantially less enrichment during the main sequence especially for slow rotators and case-3 stars have much slower surface rotation at the end of the main sequence than in the other two cases. Both of these contribute to very different evolution which should be distinguishable observationally. Indeed, whilst the models may produce similar results for full population synthesis calculations, it has been found that there is often far less agreement when different mass ranges are considered separately \citep{Brott10b}.

\subsection{Metallicity dependence}

In order to compare stellar rotation models with data it is important to distinguish which effects are observable at different metallicities. Low-metallicity stars are particularly useful because they have significantly lower mass-loss rates \citep{Vink01}. For stars of metallicity $Z=0.001$ the mass-loss rate is roughly ten times lower than at solar metallicity. This allows us to rule out effects on the models produced by our prescription for mass loss.

\begin{figure*}
\centering
\includegraphics[width=0.47\textwidth]{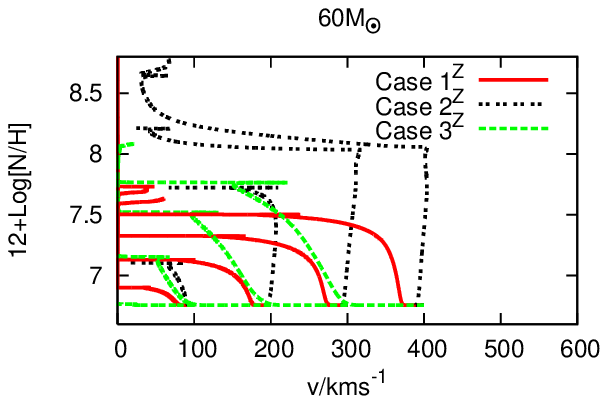}\includegraphics[width=0.47\textwidth]{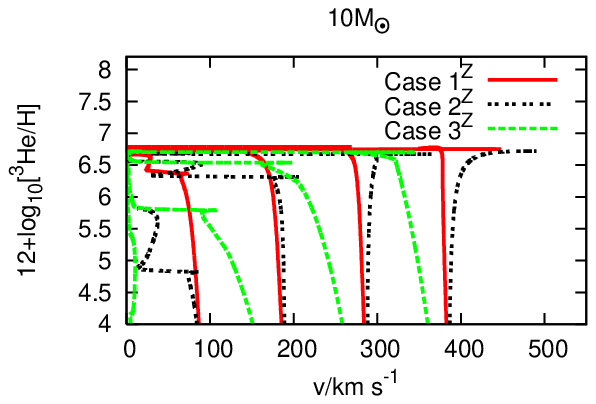}
\caption{Chemical enrichment variation at low metallicity. The left panel shows the enrichment of nitrogen and the right panel shows the enrichment of helium-3. The left-hand plot is for $60$\ms\ stars and the right-hand plot is for $10$\ms\ stars. Case~2$^Z$ now shows much higher nitrogen enrichment than case~1$^Z$, the opposite to what we found in the solar metallicity cases. The enrichment of helium-3 in case~2$^Z$ stars is still considerably less than in case~1$^Z$.}
\label{Nz1}
\end{figure*}

The low-metallicity cases show similar distinctions in the HR diagram to those at solar metallicity, although high-mass, rapidly rotating, case-2 stars are sufficiently well mixed to undergo quasi-homogeneous evolution. There are also significant differences in the nitrogen enrichment between the different models. Because the mass-loss rate is lower in low-metallicity stars they retain their surface rotation for much longer so the main sequence appears much more vertical in a Hunter diagram. Unlike the solar metallicity cases, case~2$^Z$ exhibits significantly more mixing than case~1$^Z$ particularly for slow and moderately rotating stars (Fig. \ref{Nz1}). This is the complete opposite of the results at solar metallicity and highlights the importance of testing different stellar environments to discover clues to distinguish between different models. In contrast, the enrichment of helium-3 in case~2$^Z$ stars is less than in case~1$^Z$.

\subsection{Surface gravity cut-off}

\begin{figure*}
\centering
\includegraphics[width=0.47\textwidth]{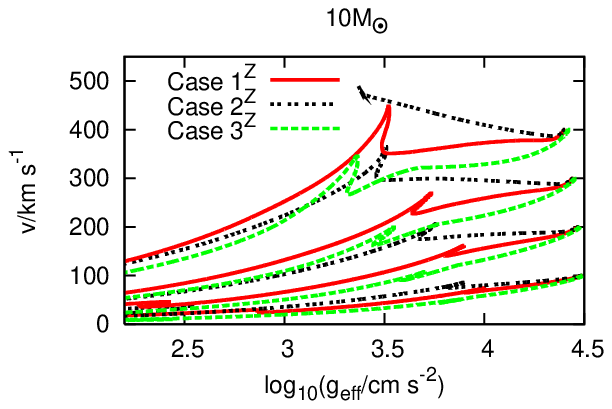}\includegraphics[width=0.47\textwidth]{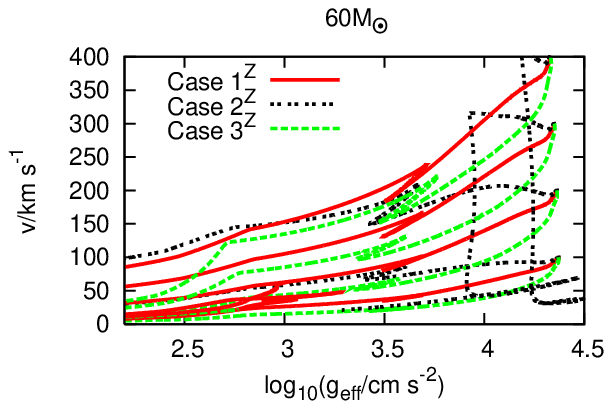}
\caption{The effective gravity variation with surface velocity for $10$\ms\ stars (left panel) and $60$\ms\ stars (right panel) for cases 1$^Z$, 2$^Z$ and 3$^Z$. The end of the main sequence occurs at the right-most cusp of each path. Although the surface gravity is similar for all the models and all initial rotation rates, the TAMS surface velocities are very different between the different cases. Exceptions to this are the rapidly-rotating, high-mass, case-2$^Z$ stars which undergo almost homogeneous evolution.}
\label{gravz}
\end{figure*}

As a consequence of increasing stellar radius and angular momentum conservation all models for stellar rotation predict a rapid decay in the surface rotation velocity after the end of the main sequence. Observations suggest that, even for rotating stars, there is a sharp cut-off in the effective gravity at $\log_{10}(g/{\rm cm^2s^{-1}})\approx 3.2$ when a star leaves the main sequence and moves over to the giant branch \citep{Brott10}. This effect depends on stars reaching the TAMS without spinning down too rapidly during the main sequence. Therefore it is more easily seen at lower metallicities where the mass-loss rate is lower. The observed value for the TAMS gravity can be enforced in rotating models by including a degree of overshooting. However this simply introduces an additional free parameter into the models. In Fig. \ref{gravz} we show the different cut-offs in the effective gravity predicted by cases 1$^Z$, 2$^Z$ and 3$^Z$. The end of the main sequence is indicated by a distinct cusp in the path of the star in the range $3<\log_{10}(g/{\rm cm^2s^{-1}})< 4$. Note that the expected number of stars with significant rotation after the main-sequence cut-off is low because the evolution to a very slowly rotating giant is extremely rapid compared to the rest of the main sequence.

For higher-mass stars there is a sharp cut-off in the surface gravity at the end of the main sequence that is the same regardless of the model used although the TAMS surface rotation is somewhat higher in case~2. The mixing in low-metallicity, high-mass, rapidly rotating, case-2 stars is very efficient and so they evolve almost homogeneously and thus they appear differently in the plot. For lower-mass stars the change in surface gravity at the end of the main sequence is still clear but varies by around an order of magnitude across all rotation rates and test cases. Case-2 and case-3 stars have lower terminal-age main-sequence surface gravities than case-1 stars and show a tendency towards lower TAMS surface gravities for more rapid rotators. Case-3 stars generally have lower rotation rates at  the end of the main sequence for low-mass stars. This distinguishes them from case~2. Once again we see that the difference in stellar properties predicted by each model is strongly dependent on rotation rate, mass and metallicity suggesting that it is essential to explore populations covering as wide a range as possible in order to test rotating models.

\subsection{Alternative models for convection}

\begin{figure*}
\centering
\includegraphics[width=0.47\textwidth]{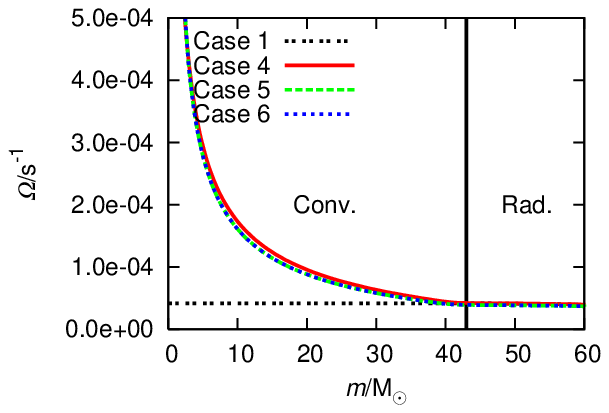} \includegraphics[width=0.47\textwidth]{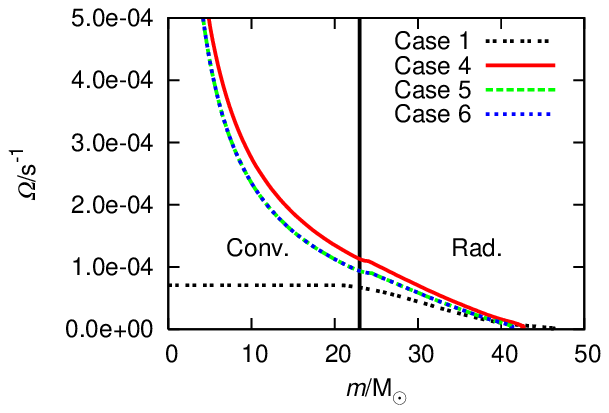}
\caption{Angular velocity distributions for $60$\ms$\,$ stars initially rotating at $300\,\rm{km}\,\rm{s}^{-1}$ for different convective models. The left panel shows the ZAMS distributions and the right panel shows the TAMS distributions. Despite there being four orders of magnitude difference between the convective diffusion coefficients in cases~4,5 and~6, there is very little difference between the angular velocity distributions they predict. In each case, the models with uniform specific angular momentum in convective zones predict more shear at the convective boundary than the models with solid body rotation in convective zones.}
\label{convdist}
\end{figure*}

In order to compare the difference in the evolution of a star owing to the details of the model for convective angular momentum transport we now focus on cases 1, 4, 5 and 6. Uniform specific angular momentum in the core causes more shear mixing near the core-envelope boundary than when the core is solid body rotating as shown in Fig. \ref{convdist}. This results in higher luminosity stars with similar temperatures. There is almost no difference between cases 4, 5 and 6 in the HR diagram.

\begin{figure*}
\centering
\includegraphics[width=0.47\textwidth]{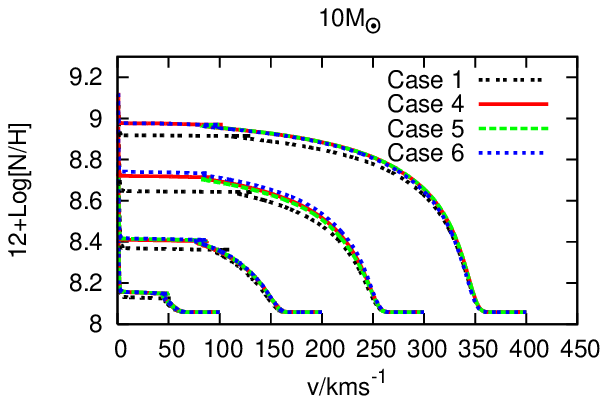} \includegraphics[width=0.47\textwidth]{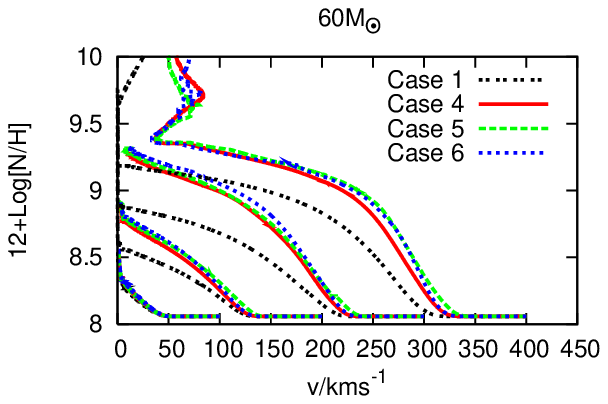}
\caption{Nitrogen enrichment variation with initial surface rotation for cases 4, 5 and 6. The left panel is for $10$\ms\ stars and the right panel is for $60$\ms\ stars. As expected the enrichment is much greater for more massive stars. There is much more enrichment for models in which the core has uniform specific angular momentum, an effect that becomes progressively more pronounced for higher masses and rotation rates}.
\label{Nsol2}
\end{figure*}

When we compare the different models for convection in a Hunter diagram we see that cases 4, 5 and 6, which have uniform specific angular momentum throughout their convective zones, have significantly more enrichment for all masses and rotation rates than case~1. The difference is more pronounced for higher-mass rapid rotators (Fig. \ref{Nsol2}). However, we note that it is more difficult to distinguish between cases 4, 5 and 6. For the highest mass stars we do find some difference in the enrichment of nitrogen and helium-3 between the models but recall that there is a difference in $D_{\rm conv}$ of four orders of magnitude between cases~5 and 6.

Given the small magnitude of the change in enrichment and structure over such a range of diffusion coefficients it seems unlikely that these tests can adequately distinguish between convective models. In addition, adjusting the calibration of case~1 could produce a very similar effect making it difficult even to distinguish between $n=0$ and $n=2$ models from observations. At the same time though, it is interesting to note the significant change that modifying the core angular momentum distribution has had on the evolution of the surface composition.

\section{Conclusions}

Rotation in stars has a number of profound effects on their evolution. Not only are there significant changes in the hydrostatic structure \citep{Endal76} but this causes a thermal imbalance that can lead to a strong meridional circulation current \citep{Sweet50}. Meridional circulation leads to additional shear which induces a number of instabilities. The resulting turbulence leads to strong mixing of both angular momentum and chemical composition.

Although most modellers include all of these effects, the exact implementation of stellar rotation can vary dramatically. For example \citet{Heger00} use a model where meridional circulation is treated diffusively. The diffusion owing to shear is the linear combination of a number of coefficients based on different possible instabilities. On the other hand, models such as those of \citet{Zahn92}, \citet{Talon97} and  \citet{Maeder03} treat the circulation as advective and use a single diffusion coefficient based on the magnitude of Kelvin-Helmholtz instabilities induced by shear. In these models it is also necessary to define the magnitude of diffusion along isobars and this too varies between different models.

We have shown that different models generally give rise to similar qualitative conclusions but there are significant differences in the results based on mass, rotation rate and metallicity. There are also open questions about how angular momentum transport occurs in convective zones.

Comparing the models based on \citet{Talon97} and \citet{Maeder03}, case~1, and that based on \citet{Heger00}, case~2, we find that case~1 gives higher luminosity stars for masses less massive than $10$\ms$\,$ and more luminous, hotter stars at higher masses than case~2. High-mass stars give similar levels of nitrogen enrichment in each case but case~2 produces far less enrichment for low-mass and intermediate-mass stars than case~1. The situation is similar for their helium-3 enrichment.

The predicted effects of rotation appear to be highly dependent on the metallicity. This is one of the clearest tests for different stellar rotation models. At metallicity of $Z=0.001$, case~2 actually produces significantly more nitrogen enrichment in high-mass stars although the degree of helium-3 enrichment is still lower in case~2 than case~1.  Additional effects are seen in lower-metallicity stars where the mass-loss rate is lower such as the variation of the surface gravity with respect to surface rotation velocity. We see a sharp cut-off in the effective gravity at the end of the main sequence but the TAMS rotation rates are very different between different models. Case-2 stars reach the end of the main sequence with higher rotation rates than the other two cases for both low and high-mass stars. Case-1 stars reach the end of the main sequence with higher surface gravity than the other two cases but only for lower-mass stars.

All current models treat convective zones as rotating solid bodies. This may be justified if convective zones can generate sufficiently strong magnetic fields \citep[e.g.][]{Spruit99} but, if not, hydrodynamic models and calculations suggest that convective zones should tend towards uniform specific angular momentum \citep{Lesaffre10,Arnett09}. Identifying whether this is the case or not is difficult from surface observations. There is no significant change in the paths of stars across the HR diagram between cases 4, 5 and 6. In fact, there is no test we have found that would adequately differentiate between these three cases  even though the diffusion coefficient in convective zones varies by four orders of magnitude. There are some minor differences in the amount of enrichment for massive stars but these would be hard to test with existing data. The models with uniform specific angular momentum generally produce slightly more luminous stars and higher surface chemical enrichment than those models with solid body rotating cores. However, slightly  less efficient mixing in the radiative zone could mask this difference.

We have thus far not included magnetic fields in the models. It has been suggested that the strong turbulence generated by rotation could result in a radiative magnetic dynamo \citep{Spruit99}. A sufficiently strong magnetic field can effectively suppress the meridional circulation and reduce the overall shear \citep{Maeder05}. It could also result in additional mass and angular momentum loss \citep[e.g.][]{Lau11}. As with rotation, there is little consensus on the details of magnetic field generation but it is generally accepted that the effects of rotation and magnetic fields cannot be considered in isolation. In the future we plan to include magnetic fields in \textsc{RoSE} in a similar manner in order to better explore this important feature of stellar evolution.

Owing to the range of available models, it is an extremely challenging problem to try to identify the one which best fits observed stellar populations. Now that large scale surveys are starting to produce data for many stars in different regions it is becoming possible to make progress and isolate which effects dominate. The key to distinguishing the relevant physics seems to be taking measurements of groups of stars at different masses, ages and metallicities. In individual clusters, models should be able to match not only the full distribution of observed stars but also the expected distribution in each mass range. Whilst most of the models can be calibrated to fit the data for a single cluster, we have shown that the behaviour of each model is highly dependent on mass and metallicity and so being able to fit data for a range of stellar environments is the true test for any model. Using our new rotating stellar evolution code \textsc{RoSE} combined with population synthesis codes \citep[e.g.][]{Izzard09,Brott10} we plan on making a detailed comparison of these models with the available data to identify how best to model stellar rotation.

\section{Acknowledgements}

The authors would like to acknowledge Raphael Hirschi for his helpful comments on the manuscript and Urs Frischknecht for compiling the data for comparison with our group's calculations. ATP thanks the STFC for his studentship and CAT thanks Churchill college for his fellowship. JJE is supported by the IoA's theory rolling grant.

\end{document}